\documentclass{iaus}
\usepackage{graphicx}

\title[Extrasolar Planetesimals] 
{The Elemental Compositions of Extrasolar Planetesimals}

\author[M. Jura]   
{M. Jura$^1$}

\affiliation{$^1$Department of Physics and Astronomy \\ University of California, Los Angeles \\
Los Angeles CA 90095-1562 USA \\ email: {\tt jura@astro.ucla.edu} }

\pubyear{2012}
\volume{293}  
\pagerange{xxx-xxx}
\jname{Formation, Detection and Characterization of Extrasolar Habitable Planets}
\editors{N.  Haghighipour, ed.}
\begin{document}

\maketitle

\begin{abstract}

Evidence is now compelling that most externally-polluted white dwarfs  
derive their heavy atoms by accretion from 
 asteroids -- the building blocks of rocky planets. Optical and ultraviolet spectroscopy of a small sample of suitable white dwarf stars shows  that to zeroth order, the accreted extrasolar parent bodies  compositionally resemble bulk Earth.  (1)  Extrasolar planetesimals are at least  85\% by mass composed of
 O, Mg, Si and Fe. (2)
Compared to the Sun, C is often deficient, usually by at least a factor of 10 and therefore comprises less than 1\% of an extrasolar planetesimal's mass.  At least to-date, C has never
been found to be enhanced as would be expected if carbon-rich planetesimals have formed.  
(3) While there may be individual exceptions, considered as a whole, the population of extrasolar asteroids accreted onto a well-defined sample of local white dwarf stars  is less than 1\% water by mass.  
\end{abstract}

\firstsection 
\section{Introduction}

The atmospheres of most white dwarf stars cooler than about 20,000 K usually are either nearly pure hydrogen or nearly pure
helium (\cite[Koester 2009]{Koeser2009}).  However, about 
25\% of these white dwarfs display heavy elements\footnote{Here, we define heavy atoms as those elements more massive than helium.}  (\cite[Zuckerman et al. 2003]{Zuckerman2003}, \cite[Zuckerman et al. 2010]{Zuckerman2010}).  Because gravitational settling
is very efficient in these compact stars, the origin of their photospheric  heavy elements must be accretion from some external source; dredge-up, radiative levitation and leftover material lingering from the white dwarf's formation can be excluded (\cite[Koester 2009]{Koester2009}).   Ten years ago, it was uncertain whether the accreted material originates in the interstellar medium
or from the star's own planetary system (\cite[Alcock et al. 1986]{Alcock1986}, \cite[Aannestad et al. 1993]{Aannestad1993}, \cite[Dupuis et al. 1993]{Dupuis1993}, \cite[Hansen \& Liebert 2003]{Hansen2003}).   
 Nowadays, as described in more detail below, amassed evidence  has provided compelling support for the scenario that the heavy elements usually derive from rocky planetesimals.  The widely accepted  model is that an asteroid's orbit is perturbed so that it  passes within the white dwarf's tidal radius where it is destroyed, a circumstellar disk is formed, and
eventually this material is accreted onto the host star  (\cite[Debes \& Sigurdsson 2002]{Debes2002}, \cite[Jura 2003]{Jura2003}).   Consequently, spectroscopic determination of the atmospheric abundances in externally-polluted white dwarfs is a  powerful
tool to measure the elemental compositions of extrasolar minor planets.

Three reasons explain why the study of externally polluted white dwarfs has progressed substantially during the past decade. 
\begin{itemize}
\item{We now have a standard model to assess; previously, there was no satisfactory scenario.  In his {\it Annual Reviews} article, ``Dusty Circumstellar Disks", 
Zuckerman (2001) wrote, ``At the moment, in spite of searches at hundreds of white dwarfs, we know of only one, G29-38, that is surrounded by dust grains .... The origin and survival of the dust is a
mystery."  While there are  open questions, current theoretical efforts are elaborations of and improvements upon the basic picture that we are witnessing the accretion of minor planets (\cite[Bonsor et al. 2011]{Bonsor2011}, \cite[Debes et al. 2012]{Debes2012a}, \cite[Metzger et al. 2012]{Metzger2012}, \cite[Rafikov 2011]{Rafikov2011}).
  }

\item{Improved infrared observational capabilities have been most valuable. In 1987, G29-38 was the first white dwarf found to display excess infrared emission (\cite[Zuckerman \& Becklin 1987]{Zuckerman1987}) produced by circumstellar dust (\cite[Graham et al. 1990]{Graham1990}). Eighteen years later, the discovery of  GD 362's infrared excess, the second white
dwarf found to have a dust disk  (\cite[Becklin et al. 2005]{Becklin2005}, \cite[Kilic et al. 2005]{Kilic2005}), opened the door to many more such detections. While further ground-based work has been useful, the {\it Spitzer Space Telescope} which was launched in 2003 greatly facilitated these efforts and about 30 white dwarfs with an infrared excess have been identified  (\cite[Farihi et al. 2009]{Farihi2009}, \cite[Farihi et al. 2012]{Farihi2012}, \cite[Brinkworth et al. 2012]{Brinkworth2012}, \cite[Xu \& Jura 2012]{Xu2012}).}

\item{High-resolution optical
spectroscopy with a 10m class telescope  have enabled detailed studies of white dwarf atmospheres.  \cite[Zuckerman et al. (2007)]{Zuckerman2007} 
reported 17 elements in the photosphere of GD 362. Abundances for all major elements -- O, Mg, Si and Fe --  were reported for the first time by  \cite[Klein et al. (2010)]{Klein2010} in the atmosphere of GD 40.  Ultraviolet spectroscopy with the {\it Cosmic Origins Spectrograph} on the {\it Hubble Space Telescope} (COS/HST) has provided excellent measurements of volatile elements such as C and S (\cite[Jura et al. 2012]{Jura2012a}, \cite[Gaensicke et al. 2012]{Gaensicke2012}, \cite[Xu et al. 2013]{Xu2013}). The study of white dwarf atmospheres provides a uniquely  powerful tool to measure the bulk compositions of extrasolar planetesimals.}
\end{itemize}

\section{Support for the Scenario that the Material Accreted onto White Dwarfs is from Rocky Planetesimals}
Observations both of polluted white dwarfs and their circumstellar disks have been important in showing that the accretion results from an ancient planetary system rather than the interstellar medium  (\cite[Jura 2008]{Jura2008}).  
\begin{itemize}
\item{Although the exact fraction is uncertain by about a factor of two (\cite[Barber et  al. 2012]{Barber2012}, \cite[Farihi et al. 2012]{Farihi2012}), approximately 2\% of white dwarfs with photospheric temperatures warmer than 10,000 K display excess infrared emission and possess circumstellar dust (\cite[Farihi et al. 2009]{Farihi2009}) with a 
characteristic temperature of ${\sim}$1000 K.  The infrared spectral energy distribution from this warm dust usually can be well-modeled by a flat, passive disk with an
inner boundary determined by the location where grains rapidly sublimate and an outer boundary determined by the tidal-radius of the white dwarf (\cite[Jura 2003]{Jura2003}, \cite[Jura et al. 2007a]{Jura2007a}). The most heavily polluted white dwarfs often display a dust disk (\cite[Kilic et al. 2006]{Kilic2006}), and all stars with dust
disks are heavily polluted (\cite[Farihi et al. 2012]{Farihi2012}).  There is no evidence for cold grains far outside the tidal radius, as would be expected if interstellar accretion is important.}
\item{Mid-infrared spectroscopy has been obtained with the {\it Infrared Spectrograph} on the {\it Spitzer Space Telescope} for eight white dwarfs with dust disks, and every 
spectrum shows  silicate emission at 10 ${\mu}$m (\cite[Reach et al. 2005]{Reach2005},
\cite[Reach et al. 2009]{Reach2009}, \cite[Jura et al. 2007b]{Jura2007b}, \cite[Jura et al 2009]{Jura2009}).  This feature is
always substantially broader than the corresponding interstellar absorption at 10 ${\mu}$m, and, instead, agrees with the emission profile seen in circumstellar silicates such as detected orbiting the main-sequence star BD +20 307 (\cite[Song et al. 2005]{Song2005}).  These data support
 scenarios where the disks derive from a circumstellar planetary system rather than from interstellar matter.}
 \item{Approximately  1/5 of all white dwarfs with dust disks also show evidence for gaseous emission, typically double peaked features in the Ca II  triplet near 8500 {\AA}.  These lines almost certainly  originate from an orbiting disk (\cite[Gaensicke et al. 2006]{Gaensicke2006}, \cite[Gaensicke et al. 2007]{Gaensicke2007}, \cite[Gaensicke et al. 2008]{Gaensicke2008}, \cite[Farihi et al. 2012]{Farihi2012}, \cite[Melis et al. 2012]{Melis2012}).    The emission line profiles  typically have full widths near 1000 km s$^{-1}$, therefore demonstrating that the gas lies within the white dwarf's tidal radius as expected in the standard model.  As best as we can tell, the dust disks and gas dusts coincide in radial extent (\cite[Melis et al. 2010]{Melils2010}, \cite[Brinkworth et al. 2012]{Brinkworth2012}).  While all white dwarfs with detected gas disks in emission also have detected dust disks,  \cite[Debes et al. (2012b)]{Debes2012b} have reported absorption from circumstellar gas orbiting WD 1124-293, an externally polluted white dwarf without a detectable dust disk.}
 \item{When measured, externally-polluted white dwarfs typically show that less than 1\% of the accreted mass is carbon (\cite[Jura 2006]{Jura2006}, \cite[Jura et al. 2012]{Jura2012a}, \cite[Gaensicke et al. 2012]{Gaensicke2012}, \cite[Koester et al. 2012]{Koester2012}).  Such a low carbon fraction is characteristic of the inner solar system (\cite[Lee et al. 2010]{Lee2010}) and completely different
 from the interstellar medium where carbon comprises more than 10\%  of the mass of elements heavier than helium.  These results strongly support the asteroidal model for the origin of the white dwarf's external pollution.}
 \item{Interstellar accretion is expected to anti-correlate
 with the star's space motion because the slower a star moves, the greater is the expected accretion rate.  However, this predicted effect is not observed.  Furthermore, a substantial number of externally-polluted white dwarfs lie outside the thin layer of Galactic gas and dust; they should not be able to accrete much interstellar matter.   Consequently, both kinematically and spatially, polluted white dwarfs do not exhibit any evidence for interstellar accretion, and, instead, the  data
 are better explained by accretion of circumstellar matter (\cite[Farihi et al. 2010]{Farihi2010}).}
 \item{The amount of mass required to account for the observed pollutions is comparable to what might be present in the asteroid belt of an extrasolar planetary system.
 The white dwarf with the largest measured  mass of pollution is SDSS J073842.56+183509.06 where the minimum amount of accreted mass is comparable to the
 mass of Ceres, the most massive asteroid in the solar system (\cite[Dufour et al. 2012]{Dufour2012}).  Beyond this extreme case, in a well-defined sample of 57 local white dwarfs with helium-dominated
 atmospheres, the  the average heavy element accretion rate has been ${\sim}$3 ${\times}$ 10$^{8}$ g s$^{-1}$ (\cite[Jura \& Xu 2012]{Jura2012b}).  Given that the average white dwarf cooling age in this sample is ${\sim}$ 2 ${\times}$ 10$^{8}$ yr (\cite[Bergeron et al. 2011]{Bergeron2011}), then each star, on average, 
accretes 2 ${\times}$  10$^{24}$ g, approximately the mass of the solar system's asteroid belt (\cite[Binzel et al. 2000]{Binzel2000}). The typical white dwarf in this sample of 57 stars is descended
 from a main-sequence star with a mass near 2.5  M$_{\odot}$.  Therefore, these stars have main-sequence lifetimes much shorter than the Sun's and their asteroid
 belt's have had much less time to lose mass  than our own solar system's.   It is plausible that
  asteroid belts orbiting these white dwarfs are substantially more massive  our solar system's  and therefore can supply the required pollution of heavy elements (\cite[Jura 2008]{Jura2008}).}
 \end{itemize}

It must be recognized that
most externally-polluted white dwarfs do not display any direct  evidence for circumstellar matter.  However, at least for stars with hydrogen-dominated atmospheres with
effective temperatures greater than 12,000 K, the gravitational settling time is much less than one year (\cite[Koester 2009]{Koester2009}).  Consequently, at least for these particular
externally-polluted white dwarfs, there is strong indirect evidence that
there is an undetected reservoir of  matter that is currently being accreted.  

The lack of measured circumstellar dust orbiting many externally polluted white dwarfs is not surprising; instead,
the detection of any orbiting circumstellar dust at ${\sim}$1000 K near even ${\sim}$10\% of all externally-polluted  white dwarfs is remarkable.  The required circular speed of ${\sim}$500 km s$^{-1}$ is so large
that two grains moving at even modest deviations from this mean motion  would mutually annihilate if they collide.  Because different asteroids characteristically have slightly different orbital angular momentum vectors, it is likely
that a disk composed of the tidal-disruption of two asteroids would produce two dust streams with sufficiently different space velocities that the end product would be largely gaseous.  Consequently, \cite[Jura (2008)]{Jura2008} argued that the observed dust disks are the result
of the destruction of one relatively massive asteroid, while gaseous disks may often be the result of multiple disruptions of smaller asteroids.   This picture allows us to understand why dust disks are associated with high accretion rates and why only a relatively few externally-polluted white dwarfs display an infrared excess from
orbiting dust.  

It is also notable, that externally-polluted white dwarfs with photospheric temperatures less  than about 9,500 K do not display dust at ${\sim}$1000 K.  The physical explanation for the lack of dust disks among cooler polluted white dwarfs is uncertain (\cite[Xu \& Jura 2012]{Xu2012}).

\section{Compositions of Extrasolar Planetesimals}

We use classical stellar spectroscopy of white dwarf atmospheres to determine the composition of the stellar photosphere and then use models to determine the composition of the polluting parent bodies.   
Because 
the derived abundances of many elements scale in approximately the same manner with  the star's effective temperature and gravity (\cite[Klein et al. 2010]{Klein2010}, \cite[Klein et al. 2011]{Klein2011}), in the best cases, uncertainties of 0.1 dex in relative abundances may  be achieved although absolute abundances may be more uncertain.
Because different heavy elements gravitationally settle with different rates, the relative abundances in a polluted white dwarf's atmosphere need not equal the
relative abundances in the accreted parent body.  \cite[Koester (2009)]{Koester2009} described three phases of an accretion event.  In the initial build-up
phase, little settling has occurred  and the relative abundances in the stellar atmosphere directly reflect the parent body composition.  In the second phase, a steady
state is achieved and the rate of accretion is balanced by the rate of gravitational settling.  In the third phase, accretion has stopped and the system ``decays"
so that the lighter elements with the longest settling time are expected to appear relatively more abundant.  Elaborations upon this simple picture can be made.
\cite[Jura \& Xu (2012)]{Jura2012}  suggested that some systems may be in a ``quasi-steady state" where there is ongoing accretion but, perhaps because
of internal disk physics (\cite[Rafikov 2011b]{Rafikov2011b}), the accretion rate is variable with time.

In the case of the white dwarfs with hydrogen dominated atmospheres that are warmer than 12,000 K, the steady state approximation almost certainly applies because
the gravitational settling times are much shorter than one year.  However, even for these stars, there can be significant uncertainties in the treatment of the atmosphere
and the derived settling times (\cite[Vennes et al. 2010]{Vennes2010}, \cite[Vennes et al. 2011]{Vennes2011}, \cite[Gaensicke et al. 2012]{Gaensicke2012}).

In the case of the white dwarfs with helium dominated atmospheres, the settling times can be as high as ${\sim}$10$^{6}$ yr.  For these stars,  the current phase of
any observed accretion event may be uncertain.  However, the white dwarfs with the largest amount of pollution are unlikely to be far advanced in any decay phase.  Also, for stars with a dust disk, likely,  there is some ongoing accretion. Therefore, at least for the most heavily polluted white dwarfs with helium dominated atmospheres,   the accretion probably either is  in a build-up phase or a  steady state or
quasi steady state phase.  For  a zeroth order measure of the compositions of extrasolar planetesimals, it does not matter which phase pertains
  (\cite[Klein et al. 2010]{Klein2010}). 
\subsection{Dominant Elements}

Earth is 94\% composed in roughly equal mass fractions of four elements: O, Mg, Si and Fe 
(\cite[Allegre et al. 2001]{Allegre2001}).  What about extrasolar planetesimals?

Detailed comparisons are best achieved for systems where all the major elements have been detected.  Currently, there are nine externally polluted white dwarfs with published detections for  O, Mg, Si and Fe.
Two especially well-studies stars with helium-dominated atmospheres are GD 40 and G241-6  where both high-resolution ground-based from Keck and ultraviolet spectra from COS/HST 
are reported 
(\cite[Jura et al. 2012]{Jura2012a}, 
\cite[Klein et al. 2010]{Klein2010}) and 13 and 12 polluting elements, respectively, are detected.   Results for these stars are shown in Figures 1-3.
If we ignore hydrogen because it may be primordial or from an earlier accretion episode,
then in the approximation that the systems are in a build-up phase, O, Mg, Si and Fe compose 92\% and 91\% of the total mass for GD 40 and G241-6, respectively (\cite[Jura et al. 2012]{Jura2012a}).     In the steady state approximation, because of their different settling times, the fractions of Fe and O increase and decrease, respectively.  However, the net change in the total mass carried by O, Mg, Si and Fe is small and these four elements still carry about
90\% of the total pollution mass (\cite[Klein et al. 2010]{Klein2010}).  
Regardless of the exact phase of the accretion episode, the four dominant elements in these two extrasolar rocky planetesimals are the same as  in bulk Earth.

The seven additional well-studied externally polluted white dwarfs where measurements of O, Mg, Si and Fe are published are HS 2253+8023 (\cite[Klein et al. 2011]{Klein2011}), PG 0843+516, PG 1015+161 and WD 1226+110 (\cite[Gaensicke et al. 2012]{Gaensicke2012}), GALEX 193156.8 +011745 (\cite[Vennes et al. 2011]{[Vennes2011}, \cite[Melis et al. 2011]{Melis2011}, \cite[Gaensicke et al. 2012]{Gaensicke2012}), SDSS J073842.56 +183509.06 (\cite[Dufour et al. 2012]{Dufour2012}), and GD 61 (\cite[Farihi et al. 2011]{Farihi2011}).  In all cases, although the  fractions for individual elements vary, the sum  of O, Mg, Si and Fe  always totals to  90\% or more of the mass  of the accreted extrasolar planetesimal.

Variations in the relative mass fractions of the four dominant elements may be signatures of the evolutionary history of the parent body.  For example, as mentioned below, GD 61 is O-enhanced and therefore the parent body possibly was ice-rich. However, at least in the case of J0738, the apparently large abundance
of O may simply reflect the uncertainties in the analysis (\cite[Dufour et al. 2012]{Dufour2012}).
 It appears that Fe is 70\% of the mass accreted onto PG 0843+516, perhaps the parent body 
resembled the planet Mercury in the solar system (\cite[Gaensicke et al. 2012]{Gaensicke2012}).  
\begin{figure}
\begin{center}
 \includegraphics[width=4.in]{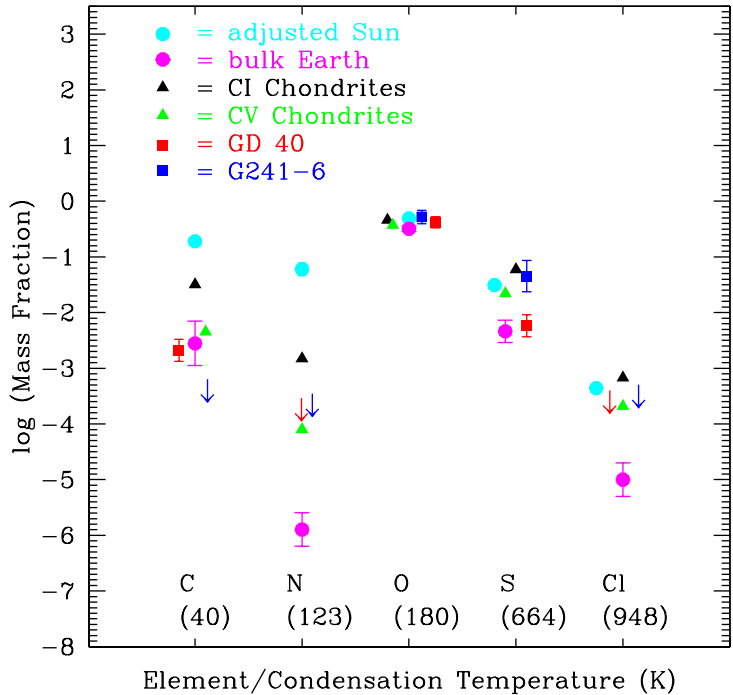} 
 \caption{ From 
 \cite[Jura et al. (2012)]{Jura2012a},
  display of  mass fractions of elements with 50\% condensation temperatures less than 1000 K  in the parent bodies accreted onto GD 40 and G241-6 and, for comparison,
 both CI and CV chondrites 
 (\cite[Wasson \& Kallemyn 1988]{Wasson1988}), bulk Earth (\cite[Allegre et al. (2001)]{Allegre2001}
  and an ``adjusted Sun" which is all elements except hydrogen, helium and the noble gases given by 
  \cite[Lodders (2003)]{Lodders2003}.  
  The temperature at which 50\% of the element condenses in the calculations of 
  \cite[Lodders (2003)]{Lodders2003} 
  is shown beneath its symbol on the plot.    We assume a build-up phase where the observed relative abundances equal the abundances in the parent body; to zeroth order these results apply to other phases of an accretion event as well (\cite[Klein et al. 2010]{Klein2010}).  We see that the composition of the planetesimals accreted onto GD 40 resembles that of bulk Earth.  Except for S, the parent body accreted onto G241-6  also has a composition 
  resembling bulk Earth's.  }
   \label{fig1}
\end{center}
\end{figure}

For stars where only some of the major elements have been detected, the percentage of mass contained in O, Mg, Si and Fe is less certain.  Available data are consistent with the view that these four elements carry at least  85\% of the mass of all studied extrasolar planetesimals.  For example, GD 362 is the white dwarf with the greatest
number of detected elements (\cite[Zuckerman et al. 2007]{Zuckerman2007}, \cite[Xu et al. 2013]{Xu2013}).  However, the effective temperature of this white dwarf with a helium-dominated
atmosphere is only 10,500 K and while Si, Mg and Fe are measured, no lines of O are detected.  To proceed, we assume that the heavy elements such as Si and Mg are carried in silicates and oxides, and therefore extrapolate to estimate the O abundance and argue that O, Si, Mg and Fe dominate.  Notably, the bulk composition of the  parent body accreted onto GD 362  resembles that of mesosiderites -- a rare class of achondritic meteorite that likely were produced from the collision of two differentiated parent bodies (\cite[Scott et al. 2001]{Scott2001}).  NLTT 43806  is another  richly polluted white dwarf where O is not detected (\cite[Zuckerman et al. 2011]{Zuckerman2011}), but, again, by assuming that
heavy elements are carried in oxide-bearing minerals,  the polluted mass is mostly carried by O, Mg, Si and Fe.   Remarkably, Al is so abundant in NLTT 43806's pollution that it is plausible that
 the accreted parent body  was the  remnant outer crustal layer of a differentiated planetesimal.

\begin{figure}
\begin{center}
 \includegraphics[width=4.in]{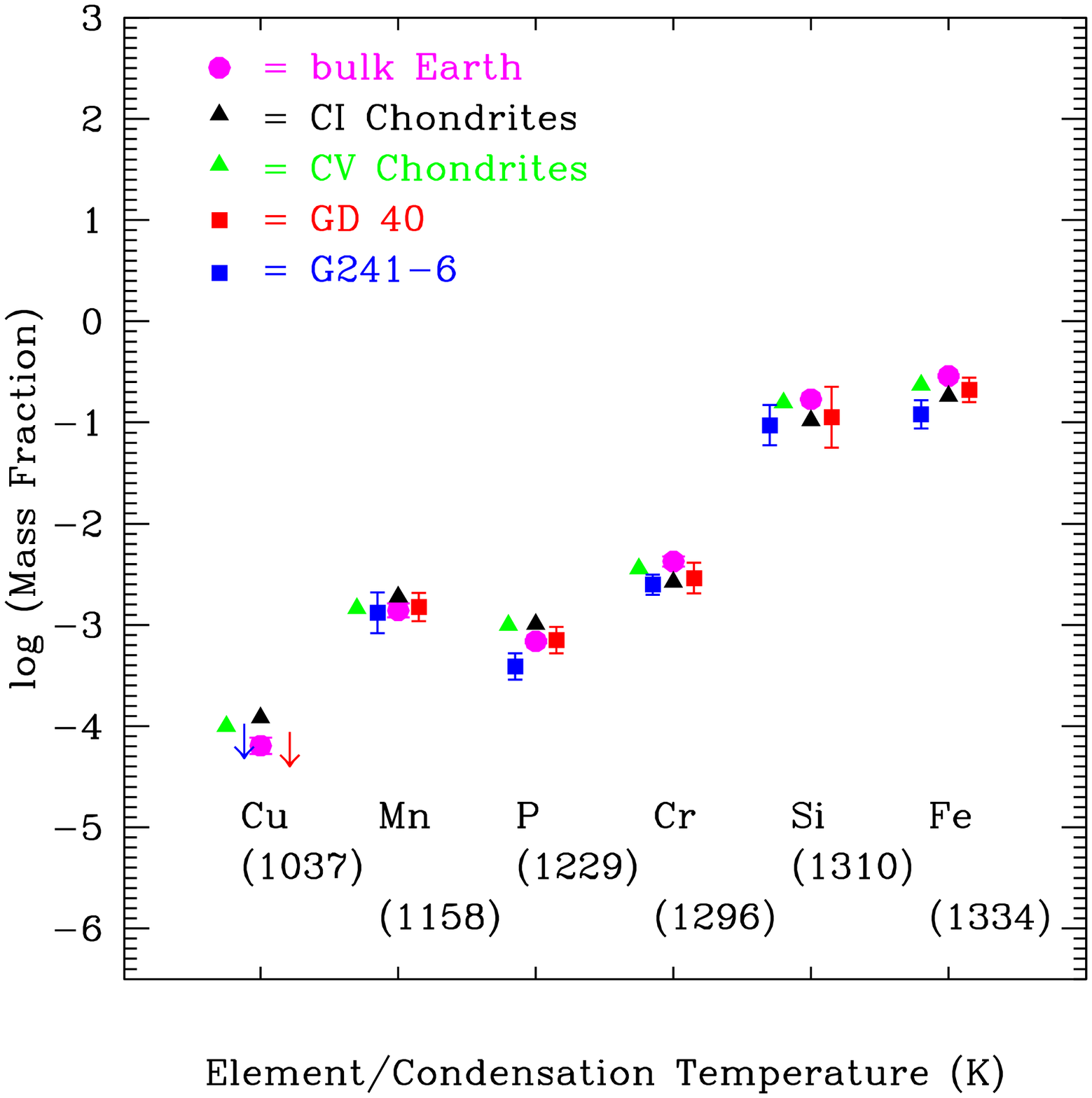} 
 \caption{Same as Figure 1 except for more refractory elements.  We do not include points for the ``corrected sun" since they are essentially equal to the values for
 CI chondrites.  The agreement in abundances between bulk Earth and the material accreted by GD 40 is striking while there are only modest differences between the abundances of bulk Earth
  and G241-6's pollution. }
   \label{fig2}
\end{center}
\end{figure}

\subsection{Carbon}

As illustrated in Figure 1, bulk Earth and all meteorites have markedly less carbon than does the Sun or the interstellar medium (\cite[Allegre et al. 2001]{Allegre2001}, \cite[Lodders 2003]{Lodder2003}).  Carbon that was contained within solid interstellar grains must have been vaporized during planetesimal formation in the inner solar system (\cite[Lee et al. 2010]{Lee2010}), and therefore, this element is a key diagnostic tracer of the evolution of rocky planetesimals.  
In almost all cases, photospheric carbon in externally-polluted white dwarfs only can be  measured in the ultraviolet.  Older, lower quality data for three
stars showed that the carbon fraction in extrasolar planetesimals was similar to Earth's and markedly lower than the Sun's (\cite[Jura 2006]{Jura2006}).  Recent, higher
quality ultraviolet observations with the {\it Far Ultraviolet Space Explorer} or COS/HST have been published for a total of 11 stars by \cite[Dupuis et al. (2007)]{Dupuis2007}, \cite[Desharnais et al. (2008)]{Desharnais2008}, \cite[Gaensicke et al. (2012)]{Gaensicke2012} and \cite[Jura et al. (2012a)]{Jura2012}.  As illustrated in Figure 1 for GD 40 and G241-6, these extrasolar planetesimals studied in detail to-date show a marked carbon deficiency compared to the solar or interstellar value.  

\cite[Koester et al. (2012)]{Koester2012} have undertaken an extensive ultraviolet survey of bright white dwarfs typically in the temperature range 17,000 K -- 25,000 K.  In their preliminary analysis,
18 stars are detected with both photospheric C and Si.  There are a few cases where the C to Si ratio approaches the solar or interstellar ratio, but for the
majority of these stars, the C to Si ratio is a factor of 10 or more lower than solar. 

The possibility that some main-sequence stars have more carbon than oxygen and therefore planet formation in their protoplanetary nebulae would lead to carbon-rich
planets has been suggested by \cite[Bond et al. (2010)]{Bond2010} and \cite[Petigura \& Marcy (2011)]{Petigura2011}.  However, it is quite uncertain whether
there are any significant number of carbon-rich main-sequence stars near the Sun (\cite[Fortney 2012]{Fortney2012}).  In any case, if planetesimals largely
composed of carbon exist, they could be found by observing polluted white dwarfs.  As yet, none has been identified.

\begin{figure}
\begin{center}
 \includegraphics[width=4.in]{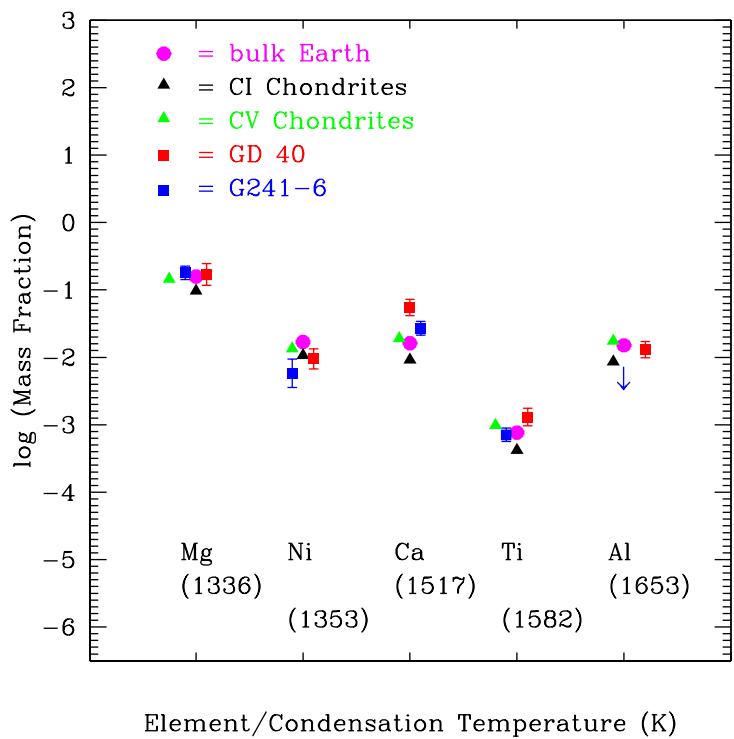} 
 \caption{ Same as Figure 2 except for even more refractory elements.  .Except for Ca, there is good agreement between the abundances of the matter accreted onto GD 40 and those in bulk Earth.}
   \label{fig3}
\end{center}
\end{figure}

\subsection{Ice}

Oxygen is so cosmically abundant that   H$_{2}$O is predicted to  be a major constituent in solid planets  that form in regions   where the temperature in the planet-forming disk  is sufficiently low that water-ice is stable.   While water is widespread in the outer
solar system (\cite[Jewitt et al. 2007]{Jewitt2007}, \cite[Encrenaz 2008]{Encrenaz2008}),  the inner solar system is relatively dry.  Most meteorites  are less than than 1\% water (\cite[Wasson \& Kallemeyn 1988]{Wasson1988}) while bulk 
Earth is probably between 0.1\% and 0.3\% water (\cite[Marty 2012]{Marty2012}).  What about extrasolar planetesimals?
 
 Because most internal water can survive within an asteroid  with a radius greater than 60 km  during a 3 M$_{\odot}$ star's pre-white-dwarf Asymptotic Giant Branch  (AGB) evolution (\cite[Jura \& Xu 2010]{Jura2010}),
measurements of  hydrogen and/or oxygen in an externally-polluted white dwarf's atmosphere  can serve as a tool to assess the amount of water in an accreted  parent body. 

 \cite[Dufour et al. (2012)]{Dufour2012} found that  there was so little H in the  material accreted onto SDSS J073842.56 +183509.06 that less than 1\% of the parent body mass was H$_{2}$O.  However, for other externally polluted white dwarfs, the upper bound to the amount of water derived by this procedure is appreciably less stringent.  
 Nevertheless, it is still possible to make a statistical estimate of the fraction of ice in an ensemble of parent bodies.
 
 The accretion of hydrogen must be treated differently from the accretion of heavy elements because hydrogen is so light that it never gravitationally settles.  Therefore, H
 accumulates over the entire cooling age of the white dwarf while the observed heavy elements only remain detectable during a gravitational settling time.
 One  approach to this complication is to consider a well defined ensemble of white dwarfs and to measure the time-averaged accretion rates of hydrogen
 and heavy elements separately.  Using this approach for  a well defined sample of 57 white dwarfs with helium-dominated atmospheres within 80 pc of the Sun, \cite[Jura \& Xu (2012)]{Jura2012a} found that water was
less than 1\% of the mass of the accreted matter in the entire ensemble of planetesimals.   While likely there are individual exceptions,  this ensemble of extrasolar
planetesimals is comparably dry to rocky objects in the inner solar system, consistent with the hypothesis that they
 formed  interior to a snow line.

Another method to study the water content in extrasolar asteroids is to measure the amount of oxygen that results from planetesimal accretion.  Is there O in excess of
what can be carried in minerals of Si and other heavy elements?  Using this approach  \cite[Farihi et al. (2011)]{Farihi2011} argued that the high fraction of oxygen in the matter accreted by GD 61 is
 a consequence of water having been abundant in the accreted object.  However, the interpretation of these data for GD 61 is not unique  (\cite[Jura \& Xu 2012]{Jura2012b}).  
\section{Conclusions}

It is now possible to make detailed measurements of the elemental abundances of extrasolar planetesimals.  In the small sample studied to date, extrasolar planetesimals are at least 85\% composed of O, Mg, Si and Fe.   Usually, the accreted planetesimals appear to be less than 1\%  C, and, as an ensemble, they are also less than 1\% water.  Future studies of externally polluted
white dwarfs will
enable a deeper understanding of the formation and evolution of extrasolar rocky planets.

\acknowledgement{This work has been partly supported by the National Science Foundation.}


\begin{thebibliography}{}

\bibitem[Aannestad et al. (1993)]{Aannestsad2003}
{Aannestad, P. A., Kenyon, S. J., Hammond, G., L., \& Sion, E. M.} 1993, \textit{AJ}, 105, 1033

\bibitem[Alcock et al. (1986)]{Alcock1986}
{Alcock, C., Fristrom, C. C., \& Siegelman, R.} 1986, \textit{ApJ}, 302, 462

\bibitem[Allegre et al. (2001)]{Allegre2001}
{Allegre, C., Manhes, G., \& Lewin, E.} 2001, \textit{Earth Planet. Sci. Lett.}, 184, 59

\bibitem[Barber et al. (2012)]{Barber2012}
{Barber, S., Patterson, A. J., Kilic, M., Leggett, S. K., Dufour, P., Bloom, J. S., \& Starr, D. L.} 2012, \textit{ApJ}, in press

\bibitem[Becklin et al. (2005)]{Becklin2005}
{Becklin, E. E., Farihi, J., Jura, M., Song, I., Weinberger, A. J., \& Zuckerman, B.} 2005, \textit{ApJ}, 632, L119

\bibitem[Bergeron et al. (2011)]{Bergeron2011}
{Bergeron, P., Wesenael, F., \& Dufour, P. et al.} 2011, \textit{ApJ}, 737, 28

\bibitem[Binzel et al. (2000)]{Binzel2000}
{Binzel, R. P., Hanner, M. S., \& Steel, D. I.} 2000, in \textit{Allen's Astrophysical Quantities}, A. N. Cox, ed., 315

\bibitem[Bond et al. (2010)]{Bond2010}
{Bond, J. C., O'Brien, D. P., \& Lauretta, D. S.} 2010, \textit{ApJ}, 715, 1050

\bibitem[Bonsor et al. (2011)]{Bonsor2011}
{Bonsor, A., Mustill, A. J., \& Wyatt, M. C.} 2011, \textit{MNRAS}, 414, 930

\bibitem[Brinkworth et al. (2012)]{Brinkworth2012}
{Brinkworth, C., Gaensicke, B., Girven, J. M., Hoard, D. W., Marsh, T. R., Parsons, S. G., \& Koester, D.} 2012, \textit{ApJ}, 750, 86

\bibitem[Debes \& Sigurdsson (2002)]{Debes2002}
{Debes, J. H., \& Sigurdsson, S.} 2002, \textit{ApJ}, 572, 556

\bibitem[Debes et al. (2012a)]{Debes2012a}
{Debes, J. H., Walsh, K. J., \& Stark, C.} 2012a, \textit{ApJ} 747, 148

\bibitem[Debes et al. (2012b)]{Debes2012b}
{Debes, J. H., Kilic, M., Faedi, F., Shkolnik, E. L., Lopez-Morales, M., Weinerger, A. J., Slesnick, C., \&  West, R. G.} 2012b, \textit{ApJ}, 754, 59

\bibitem[Desharnais et al. (2008]{Desharnais2008}
{Desharnais, S., Wesemael, F., Chayer, P., Kruk, J. W., \& Saffer, R. A.} 2008, \textit{apJ}, 672, 540

\bibitem[Dufour et al. (2012)]{Dufour2012}
{Dufour, P., Kilic, M., fontaine, G., Bergeron, P., Melis, C., \& Bochanski, J.} 2012, \textit{ApJ}, 749, 6

\bibitem[Dupuis et al. (1993)]{Dupuis1993}
{Dupuis, J.., Fontaine, G., \& Wesemael, F.} 1993, \textit{ApJS}, 87, 345

\bibitem[Dupuis et al. (2007)]{Dupuis2007}
{Dupuis, J., Bouabid, M.-P., Wesemael, F., \& Chayer, B.} 2007, in \textit{15th European Workshop on White Dwarfs, ASP Conference Series}, eds. R. Napiwotzki \& M. R. Burleigh, 372, 261

\bibitem[Encrenaz (2008)]{Encrenaz2008}
{Encrenaz, T.} 2008, \textit{ARAA}, 46, 57

\bibitem[Farihi et al. (2009)]{Farihi2009}
{Farihi, J., Jura, M., \& Zuckerman, B.} 2009, \textit{ApJ}, 694, 805

\bibitem[Farihi et al. (2010)]{Farihi2010}
{Farihi, J., Barstow, M. A., Redfield, S., Dufour, P., \& Hambly, N. C.} 2010, \textit{MNRAS}, 404, 2123

\bibitem[Farihi et al. (2011)]{Farihi2011}
{Farihi, J., Brinkworth, C. S., Gaensicke, B. T. et al.} 2011, \textit{ApJ}, 728, 8

\bibitem[Farihi et al. (2012)]{Farihi2012}
{Farihi, J., Gaensicke,  B. T., Steele, P. R., Girven, J.,Burleigh, M. R., Breedt, E., \& Koester, D.} 2012, \textit{MNRAS}, 421, 1635

\bibitem[Fortney (2012)]{Fortney2012}
{Fortney, J.} 2012, \textit{ApJ}, 747, L27

\bibitem[Gaensicke et al. (2006)]{Gaensicke2006}
{Gaensicke, B. T., Marsh, T. R., Southworth, J., \& Rebassa-Mansergas, A.} 2006, \textit{Science}, 314, 1908

\bibitem[Gaensicke et al. (2007)]{Gaensicke2007}
{Gaensicke, B. T., Marsh, T. R., \& Southworth, J.} 2007, \textit{MNRAS}, 380, L35

\bibitem[Gaensicke et al. (2008)]{Gaensicke2008}
{Gaensicke, B. T., Koester, D., Marsh, T. R., Rebassa-Mansergas, A. \& Southworth, J.} 2008, \textit{MNRAS}, 391, L103

\bibitem[Gaensicke et al. (2012)]{Gaensicke2012}
{Gaensicke, B. T., Koester, D., Farihi,J., Girven, J., Parson, S. G., \& Breedt, E.} 2012, \textit{MNRAS}, 424, 333

\bibitem[Graham et al. (1990)]{Graham1990}
{Graham, J. R., Matthews, K., Neugebauer, G., \& Soifer, B. T.} 1990, \textit{ApJ}, 357, 216

\bibitem[Hansen \& Liebert (2003)]{Hansen2003}
{Hansen, B. M. S., \& Liebert, J.} 2003, \textit{ARAA}, 41, 465

\bibitem[Jewitt et al. (2007)]{Jewitt2007}
{Jewitt, D., Chizmadia, L., Grimm, R., \& Prialnik, D.} 2007, in \textit{Protostars and Planets V}, B. Reipurth, D. Jewitt \& K. Keil, eds., (University of Arizona Press: Tucson), 863

\bibitem[Jura (2003)]{Jura2003}
{Jura, M.} 2003, \textit{ApJ}, 584, L91

\bibitem[Jura (2006)]{Jura2006}
{Jura, M.} 2006, \textit{ApJ}, 653, 613

\bibitem[Jura (2008)]{Jura2008}
{Jura, M.} 2008, \textit{AJ}, 135, 1785

\bibitem[Jura et al. (2007a)]{Jura2007a}
{Jura, M., Farihi, J., \& Zuckerman, B.} 2007a, \textit{ApJ}, 663, 1285

\bibitem[Jura et al. (2007b)]{Jura2007b}
{Jura, M., Farihi, J., Zuckerman, B., \& Becklin, E. E.} 2007b, \textit{AJ}, 133, 1927

\bibitem[Jura et al. (2009)]{Jura2009}
{Jura, M., Farihi, J., \& Zuckerman, B.} 2009, \textit{AJ},  137, 3191

\bibitem[Jura et al. (2012a)]{Jura2012a}
{Jura, M., Xu, S., Klein, B., Koester, D., \& Zuckerman, B.} 2012a, \textit{ApJ}, 750, 69

\bibitem[Jura \& Xu (2010)]{Jura2010}
{Jura, M., \& Xu, S.} 2010, \textit{AJ}, 143, 6

\bibitem[Jura \& Xu (2012)]{Jura2012b}
{Jura, M. \& Xu, S.} 2012, \textit{AJ}, 143, 6

\bibitem[Kilic et al. (2005)]{Kilic2005}
{Kilic, M., von Hippel, T., Leggett, S. K., \& Winget, D. E.} 2005, \textit{ApJ}, 632, L115

\bibitem[Kilic et al. (2006)]{Kilic2006}
{Kilic, M., von Hippel, T., Leggett, S. K., \& Winget, D. E.} 2006, \textit{ApJ}, 646, 474

\bibitem[Klein et al. (2010)]{Klein2010}
{Klein, B., Jura, M., Koester, D., Zuckerman, B., \& Melis, C.} 2010, \textit{ApJ}, 709, 950

\bibitem[Klein et al. (2011)]{Klein2011}
{Klein, B., Jura, M., Koester, D., \& Zuckerman, B.} 2011, \textit{ApJ}, 741, 64

\bibitem[Koester (2009)]{Koester2009}
{Koester, D.} 2009, \textit{A}\&\textit{Ap}, 498, 517

\bibitem[Koester et al. (2012)]{Koester2012}
{Koester, D., Gaensicke, B., Girven, J., \& Farihi, J.} 2012, in {\it 18th European White Dwarf Conference}, in press

\bibitem[Lee et al. (2010)]{Lee2010}
{Lee, J.-E., Bergin, E. A., \& Nomura, H.} 2010, \textit{ApJ}, 710, L21

\bibitem[Lodders (2003)]{Lodders2003}
{Lodders, K.} 2003, \textit{ApJ}, 591, 1220

\bibitem[Marty (2012)]{Marty2012}
{Marty, B.} 2012, \textit{Earth and Planet. Sci. Lett.}, 313, 56

\bibitem[Melis et al. (2010)]{Melis2010}
{Melis, C., Jura, M., Albert, L., Klein, B., \& Zuckerman, B.} 2010, \textit{ApJ}, 724, 470

\bibitem[Melis et al. (2011)]{Melis2011}
{Melis, C., Farihi, J., Dufour, P. et al.} 2011, \textit{ApJ}, 732, 90

\bibitem[Melis et al. (2012)]{Melis2012}
{Melis, C., Dufour, P., Farihi, J. et al.} 2012, \textit{ApJ}, 751, 4

\bibitem[Metzger et al. (2012)]{Metzger2012}
{Metzger, B. D., Rafikov, R. R., \& Bochkarev, K. V.} 2012, \textit{MNRAS}, 423, 505

\bibitem[Petigura \& Marcy (2011)]{Petigura2011}
{Petigura, E. \& Marcy, G. W.} 2011, \textit{ApJ}, 735, 41

\bibitem[Rafikov (2011a)]{Rafikov2011a}
{Rafikov, R.} 2011a, \textit{ApJ}, 732, L3

\bibitem[Rafikov (2011b)]{Rafikov2011b}
{Rafikov, R.} 2011b, \textit{MNRAS}, 416, L55

\bibitem[Reach et al. (2005)]{Reach2005}
{Reach, W. T., Kuchner, M. J., von Hippel, T., Burrows, A., Mullally, F., Kilic, M., \& Winget, D. E.} 2005, \textit{ApJ}, 635, L161

\bibitem[Reach et al. (2009)]{Reach2009}
{Reach, W. T., Lissey, C., von Hippel, T., \& Mullally, F.} 2009, \textit{ApJ}, 693, 697

\bibitem[Scott et al. (2001)]{Scott2011}
{Scott, E. R. D., Haack, H., \& Love, S. G.} 2001, \textit{Meteoritics \& Planetary Sci.}, 36, 869

\bibitem[Song et al. (2005)]{Song2005}
{Song, I., Zuckerman, B., Weinberger, A J., \& Becklin, E. E.} 2005, Nature, 436, 363

\bibitem[Vennes et al. (2010)]{Vennes2010}
{Vennes, S., Kawka, A., \& Nemeth, P.} 2010, \textit{MNRAS}, 404, L40

\bibitem[Vennes et al. (2011)]{Vennes2011}
{Vennes, S., Kawka, A., \& Nemeth, P.} 2011, \textit{MNRAS}, 413, 2545

\bibitem[Wasson \& Kallemeyn (1988)]{Wasson1988}
{Wasson, J. D. \& Kallemeyn, G.W.} 1988, \textit{Phil. Trans. R. Soc. A}, 325, 535

\bibitem[Xu \& Jura (2012)]{Xu2012}
{Xu, S. \& Jura, M.} 2012, \textit{ApJ}, 745, 88

\bibitem[Xu et al. (2013)]{Xu2013}
{Xu, S., Jura, M., Klein, B., Koester, D., \& Zuckerman, B.} 2013, submitted

\bibitem[Zuckerman \& Becklin (1987)]{Zuckerman2003}
{Zuckerman, B., \& Becklin, E. E.} 1987, Nature, 330, 138

\bibitem[Zuckerman (2001)]{Zuckerman2001}
{Zuckerman, B.} 2001, \textit{ARAA}, 39, 549

\bibitem[Zuckerman et al. (2003)]{Zuckerman2003}
{Zuckerman, B., Koester, D., Reid, I. N., \& Hunsch, M.} 2003, \textit{ApJ}, 596, 477

\bibitem[Zuckerman et al. (2007)]{Zuckerman200}
{Zuckerman, B. Koester, D., Melis, C., Hansen, B., \& Jura, M.} 2007, \textit{ApJ}, 671, 872

\bibitem[Zuckerman et al. (2010)]{Zuckerman2010}
{Zuckerman, B., Melis, C., Klein, B., Koester, D., \& Jura, M.} 2010, \textit{ApJ}, 722, 725

\bibitem[Zuckerman et al. (2011)]{Zuckerman2011}
{Zuckerman, B., Koester, D., Dufour, P., Melis, C., Klein, B., \& Jura, M.} 2011, \textit{ApJ}, 739, 101

\end{thebibliography}
\end{document}